\documentclass[11pt]{article}
\usepackage{mainstyle}
\usepackage[utf8]{inputenc}
\usepackage[noend]{algpseudocode}
\usepackage[backend=bibtex, maxnames=4, sorting=none, citestyle=numeric-comp, bibstyle=trad-unsrt]{biblatex}
\usepackage{soul}

\usepackage[a4paper,
            left=2cm,
            right=2cm,
            top=2cm,
            bottom=2cm]{geometry}

\title{A numerical framework for Newtonian-noise estimation at the Einstein Telescope: \\ 2-D simulations beyond the plane-wave approximation}

\author{Patrick Schillings $^{1\,*}$, Shi Yao$^{2,3\,*}$, Johannes Erdmann$^1$, Andreas Rietbrock$^2$}

\date{\small$^1$ RWTH Aachen University, III. Physikalisches Institut A, Aachen, Germany\\
\small$^2$ Karlsruhe Institute of Technology, Geophysical Institute, Karlsruhe, Germany\\
\small$^3$ Technische Universität Dresden, Institute for Nuclear and Particle Physics (IKTP), Dresden, Germany\\
\small $^*$ Both authors  contributed equally.\\
\small patrick.schillings@rwth-aachen.de\\
\small shi.yao@partner.kit.edu}

\begin{document}

\maketitle

\begin{abstract}
\noindent
The Einstein Telescope (ET) is a third-generation underground gravitational-wave observatory designed to extend the detection sensitivity down to a few Hertz. Newtonian noise is expected to limit the low-frequency sensitivity of ET, particularly in the 3-15~Hz band. Most existing estimates rely on analytical or semi-analytical models assuming homogeneous or layered media, neglecting geological heterogeneity and complex wave interactions. In this work, we present a numerical framework for Newtonian-noise estimation based on spectral-element simulations of a seismic wave field. As a proof of concept, we first benchmark the numerical results against analytical plane-wave predictions in a two-dimensional homogeneous medium with a single surface source, demonstrating excellent agreement for both bulk and cavern contributions. We then extend the model to an array of 30 stochastic surface sources to approximate stationary ambient seismic excitation. The P-wave fraction inferred from the simulated wave field is, in this simple homogeneous case, significantly lower than commonly assumed, indicating enhanced prospects for Newtonian-noise mitigation. The framework is readily applicable to three-dimensional simulations and to integration of detailed local seismic models and topography, offering strong potential for site-specific Newtonian-noise estimation.\\

\end{abstract}

\section{Introduction}
The Einstein Telescope (ET) is a planned European underground gravitational wave detector that is designed to largely improve the sensitivity of gravitational wave measurements~\cite{ETDesignReportUpdate}. With two jointly implemented laser interferometers, each optimized for a specific frequency band, the frequency range is extended down to 3~Hz and up to several kHz.
The sensitivity of the low frequency interferometer is limited by Newtonian noise~\cite{NewtonianNoiseOrigin} in the frequency range between 3 and 15~Hz. 
Newtonian noise in an underground detector is mainly caused by seismic body waves~\cite{SiteSelectionCriteria}:
P-waves, which are longitudinal and lead to density fluctuations in the rock and at boundaries between geological layers of different density, most importantly the cavern walls, and S-waves, which are transversal and lead to density fluctuations only at such boundaries. In both cases, for P- and for S-waves, this leads to a gravitational effect on the ET test masses, known as Newtonian noise.

Estimates for Newtonian noise have been studied for the Einstein telescope~\cite{HildsNoiseEstimation,SiteSelectionCriteria,terrestialGravityFluctuations,SosESiteChara,EMRSiteChara,lowerLimit,janssens2022impact, janssens2024correlated} as well as techniques for its mitigation~\cite{terrestialGravityFluctuations,CellaPredictionOfNN,vanBeveren:2023seq,FrancescaSingleMirrorOptimization,FrancescaJointMirrorOptimization,MyFirstPaper, rading2025distributed, ophardt2025silencing}.
However, in these approaches, the seismic field is often approximated as a plane wave, where the displacement at a given location is derived from analytical expressions assuming a homogeneous medium and infinite coherence~\cite{HildsNoiseEstimation,SiteSelectionCriteria,terrestialGravityFluctuations,SosESiteChara,EMRSiteChara,lowerLimit,janssens2022impact, janssens2024correlated,FrancescaSingleMirrorOptimization,FrancescaJointMirrorOptimization,ophardt2025silencing,MyFirstPaper}, or from semi-analytical models that assume horizontal layering and lateral homogeneity~\cite{vanBeveren:2023seq}. The Newtonian noise is then derived from the computed or measured displacement field~\cite{terrestialGravityFluctuations}. Although this captures first-order behaviour, it neglects the effects of geological heterogeneity, like scattering, mode conversions and finite geometries, as well as correlations between the different wave types and anisotropies in the wave field.
Based on analytical solutions some studies~\cite{FrancescaSingleMirrorOptimization,FrancescaJointMirrorOptimization,MyFirstPaper,ophardt2025silencing} suggest that by using seismic array measurements the Newtonian noise at the mirror location can be predicted. This would be a valid approach if the Newtonian noise can be predicted from the displacement field at a single point.
Acquiring sufficiently dense seismic array data is challenging in practice, particularly for underground installations. To overcome these limitations, Newtonian-noise estimation should extend beyond analytical approximations and integrate fully numerical wave-field simulations capable of resolving geological heterogeneity and complex wave interactions.

In this paper, we show a first proof of concept towards this goal by comparing the simulated results to the analytical ones in a simple homogeneous case with a single source. We then extend this scenario to an array of sources and try a first estimate of the relative P- and S-wave content of such a wave field. Section~\ref{sec:NN} presents the analytical description of Newtonian noise and introduces the numerical simulation that has been used. Section~\ref{sec:results} shows the results: a cross check with the analytical calculations for a single source and an estimation of the noise level for an array of 30~sources. For the latter, we also provide an estimate of the P-wave fraction in the Newtonian-noise estimate, which is an important parameter for its mitigation potential~\cite{terrestialGravityFluctuations}.
Section~\ref{sec:conclusions} provides a summary and an outlook.

\section{Newtonian-noise estimation}
\label{sec:NN}

\subsection{Analytical description}
\label{sec:analytical}

\label{sec:NN_theo}
Newtonian noise is the gravitational force exerted by density fluctuations $\delta \rho$ due to the seismic wave displacement field $\vec\xi$, both of which depend on location $\vec x$ and time $t$. These density fluctuations can be derived from the continuity equation for seismic events in rock~\cite{terrestialGravityFluctuations} (simplified for $\delta\rho\ll\rho$):
\begin{equation}
    \delta\rho(\vec{x},t)=-\vec\nabla\bracket{\rho(\vec{x})\vec\xi(\vec{x},t)} ,
    \label{eq:NN_continuity_eq}
\end{equation}

where $\rho$ is the stationary density field without any displacement effects.

It can be decomposed into a bulk contribution, arising from regions where the background density is constant \((\rho(\vec{x})=\rho)\), and a surface contribution, arising at interfaces between different media where the stationary density field \(\rho(\vec{x})\) changes abruptly:
\begin{equation}
        \delta\rho(\vec{x},t)=-\rho\vec\nabla\cdot\vec\xi(\vec{x},t)-\vec\xi(\vec x,t)\cdot\vec \nabla\rho(\vec x) .
        \label{eq:continuity}
\end{equation}

For the second term, smooth density gradients with depth are neglected, since they produce only slowly varying contributions over the upper few kilometers and do not significantly affect the resulting Newtonian noise. The dominant density contrasts are associated with discrete geological interfaces, including sedimentary layer boundaries, fault zones, and cavern walls.

The gravitational force fluctuations $\delta \vec F_M$ at a test-mass location $\vec x_M$ with mass $M$ originating from these density fluctuations are given by:
\begin{equation}
    \delta \vec F_M(t)=GM\int \dd\vec x \frac{\delta\rho(\vec x,t)}{\abs{\vec x-\vec x_M}^3}\bracket{{\vec x-\vec x_M}} ,
    \label{eq:NewtonsBasic}
\end{equation}
where $G$ is the gravitational constant.

The first term of Eq.~\eqref{eq:continuity} describes density fluctuations in the bulk medium. When inserted into Eq.~\eqref{eq:NewtonsBasic}, this term gives the bulk contribution:

\begin{equation}
    \delta\vec F_{M,\text{bulk}}(t)=-GM\rho\int \dd\vec x \frac{\vec \nabla\cdot\vec\xi(\vec x,t)}{\abs{\vec x-\vec x_M}^3}\bracket{{\vec x-\vec x_M}} .
    \label{eq:bulk_integral}
\end{equation}

In the same way, the second term of Eq.~\eqref{eq:continuity} can be rewritten at a surface $\vec\Sigma(\vec x)$ with $\rho(\vec x)=0$ (neglecting the finite density of air) on one side and $\rho(\vec x)=\rho$ on the other side with $\vec\nabla\rho(\vec x)=-\rho \vec{n}(\vec x)\delta(\vec \Sigma(\vec x)-\vec{x})$ and inserted into Eq.~\eqref{eq:NewtonsBasic})\:
\begin{equation}
    \delta \vec F_{M,\text{surf}}(t)=GM\rho\int \dd\vec \Sigma \,\frac{\vec n(\vec x)\cdot\vec\xi(\vec x,t)}{\abs{\vec x-\vec x_M}^3}\bracket{{\vec x-\vec x_M}} ,
    \label{eq:surface_integral}
\end{equation}
where $\vec n$ is the normal surface vector pointing to the \say{outside}. To generalize this to arbitrary interfaces, one can replace $\rho$ with the density difference $\Delta\rho$. This is known as the \say{surface integral}.

The result of the bulk integral for a monochromatic plane P-wave is
\begin{equation}
    \delta \vec F_\text{bulk}(\vec x_M,t)= \frac{12\pi}{3}GM\rho\cdot \vec \xi(\vec x_M,t) ,
    \label{eq:monoBulkResult}
\end{equation}
where $\vec\xi(\vec x_M,t)$ is the (virtual) displacement field at the mirror position.

The result of the surface integral over a small spherical cavern, evaluated in the centre of this cavern, for P-waves as well as for S-waves, is
\begin{equation}
    \delta\vec F_\text{cavern}(\vec x_M,t)= -\frac{4\pi}{3}GM\rho\cdot \vec \xi(\vec x_M,t) .
    \label{eq:monoCavernResult}
\end{equation}

The bulk and surface integral can be summarized by partial integration into the following equation (\say{dipole equation}). This equation contains both, bulk and surface terms for a given displacement field $\rho(\vec x)$:
\begin{equation}
        \delta \vec F_{M}(t)=GM\int \dd\vec x \frac{\rho(\vec x)}{\abs{\vec x-\vec x_M}^3}\bracket{ \vec \xi(\vec x,t) -3\bracket{\frac{\vec x-\vec x_M}{\abs{\vec x-\vec x_M}}\cdot\vec \xi(\vec x,t)}\frac{\vec x-\vec x_M}{\abs{\vec x-\vec x_M}}} .
        \label{eq:dipole}
\end{equation}

This formulation includes virtual Newtonian-noise contributions from the outer boundary if the integration domain is not chosen large enough. To reach high precision, one should integrate over multiple wavelengths of the relevant seismic waves, but as an estimation, lower integration limits (of the order of one wavelength) are sufficient.

\subsection{Numerical framework}
\label{sec:sim}

\begin{figure}[htbp]
    \centering
    \includegraphics[width=0.5\linewidth]{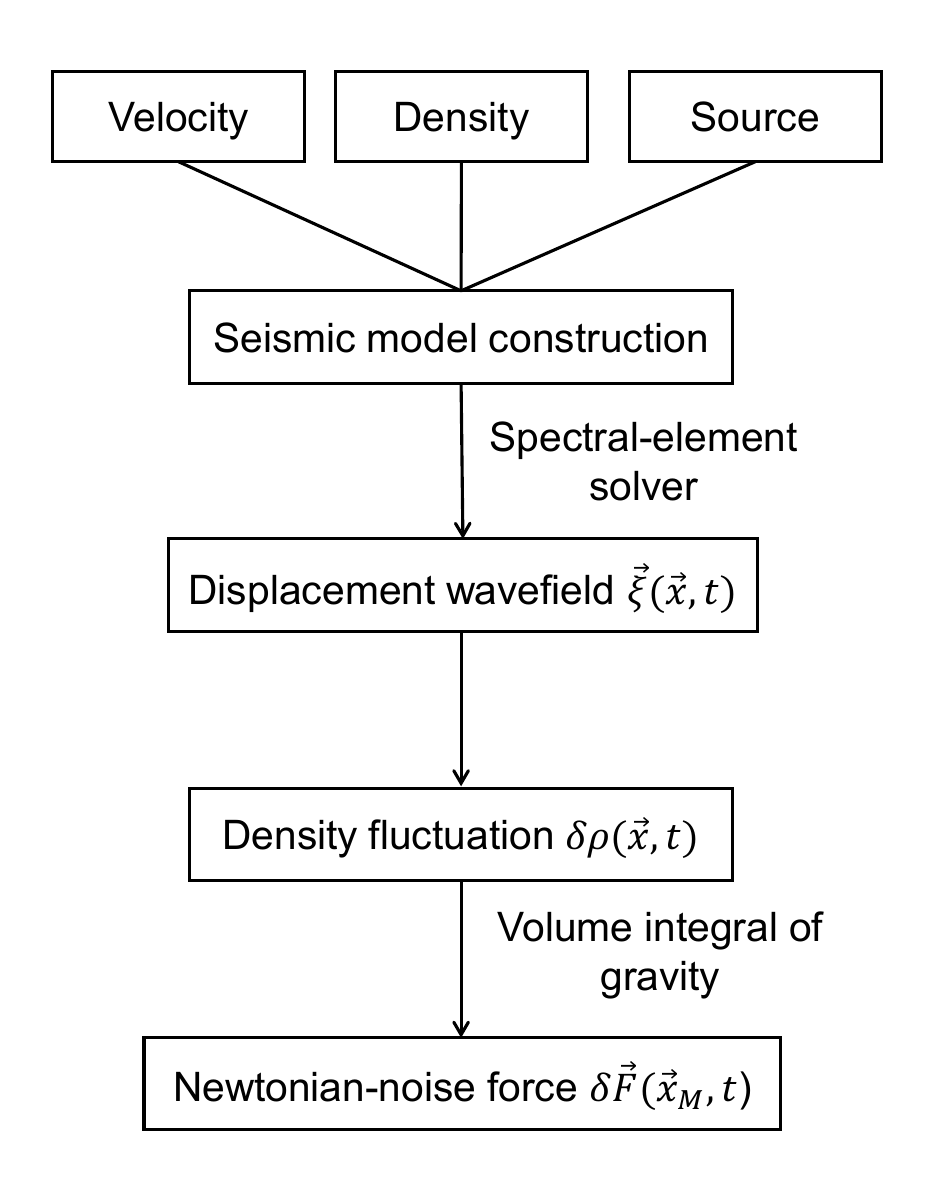}
    \caption{Workflow for the numerical estimation of Newtonian noise from simulated seismic wave fields. The volume integral refers to Eq.~\eqref{eq:NewtonsBasic}, which we normally use in the form of Eq.~\eqref{eq:dipole}. Only for validation, we also use the bulk formulation Eq.~\eqref{eq:bulk_integral}.}
    \label{fig:workflow}
\end{figure}

The overall workflow for the numerical framework is illustrated in Fig.~\ref{fig:workflow}. We begin by constructing a seismic model defined by the spatial distributions of velocity and density, together with a customized seismic source.

The seismic wave field can then be obtained by solving the seismic wave equation, which governs the evolution of ground motion in response to geological structure and external forces.
In a continuum medium, the governing equation of motion follows from conservation of momentum and can be written as:
\begin{equation}
\rho \, \frac{\partial^2 \vec{\xi}(\vec{x},t)}{\partial t^2}
=
\vec\nabla \cdot \vec{\sigma}(\vec{x},t)
+
\vec{f}(\vec{x},t),
\label{eq:seismic}
\end{equation}

where $\rho$ is the density, $\vec{\xi}$ is the displacement vector, $\vec{\sigma}$ is the stress tensor, and $\vec{f}$ represents external force densities.

In this study, the seismic wave equation is solved using Salvus, a highly parallelized and GPU-accelerated spectral-element solver~\cite{afanasiev2019salvus}. The spectral-element method integrates the geometric adaptability of finite-element formulations with high-order polynomial basis functions characteristic of spectral methods~\cite{komatitsch1999gji}, thereby achieving both numerical accuracy and computational efficiency in heterogeneous media~\cite{afanasiev2019salvus}. Based on the displacement wave field obtained from the numerical simulations, the corresponding density fluctuations can be computed using Eq.~\eqref{eq:NN_continuity_eq}.

Once the density perturbation field is computed, its gravitational contribution at a given position $\vec x_M$ is evaluated by integrating over the relevant volume to compute the resulting Newtonian-noise force (see Eq.~\eqref{eq:NewtonsBasic}).

\section{Results}
\label{sec:results}

Two numerical cases are considered, where we limit ourselves to two-dimensional examples for simplicity in this proof-of-concept study: we first examine the wave field of a single source at the surface above the test mass as a simple benchmark to validate the numerical framework under controlled conditions; we then study an array of 30 simultaneous sources at the surface to approximate stationary ambient seismic noise.

Both cases are based on the same homogeneous medium defined in a domain of $20\,\mathrm{km} \times 8\,\mathrm{km}$. The top boundary is treated as a free surface, while absorbing boundary layers are applied along the remaining three boundaries to suppress artificial reflections. The elastic parameters, the compressional-wave velocity $c_P = \SI{3000}{\meter/\second}$, the S-wave velocity $c_S = c_P/1.7$, and the density $\rho$, computed from the empirical $c_P$-$\rho$ relation~\cite{brocher2005empirical}, is approximately $\SI{2224} {\kilogram/\cubic\meter}$, are selected to represent a mechanically consistent homogeneous upper-crustal reference medium. This parameter set yields a physically admissible isotropic elastic model with positive Lamé parameters and a realistic $c_P/c_S$ ratio ($\approx 1.70$), ensuring stable P-S-wave separation and well-defined impedance.

\subsection{A single source in a homogeneous medium}

\begin{figure}[htbp]
    \centering
    \includegraphics[width=\linewidth]{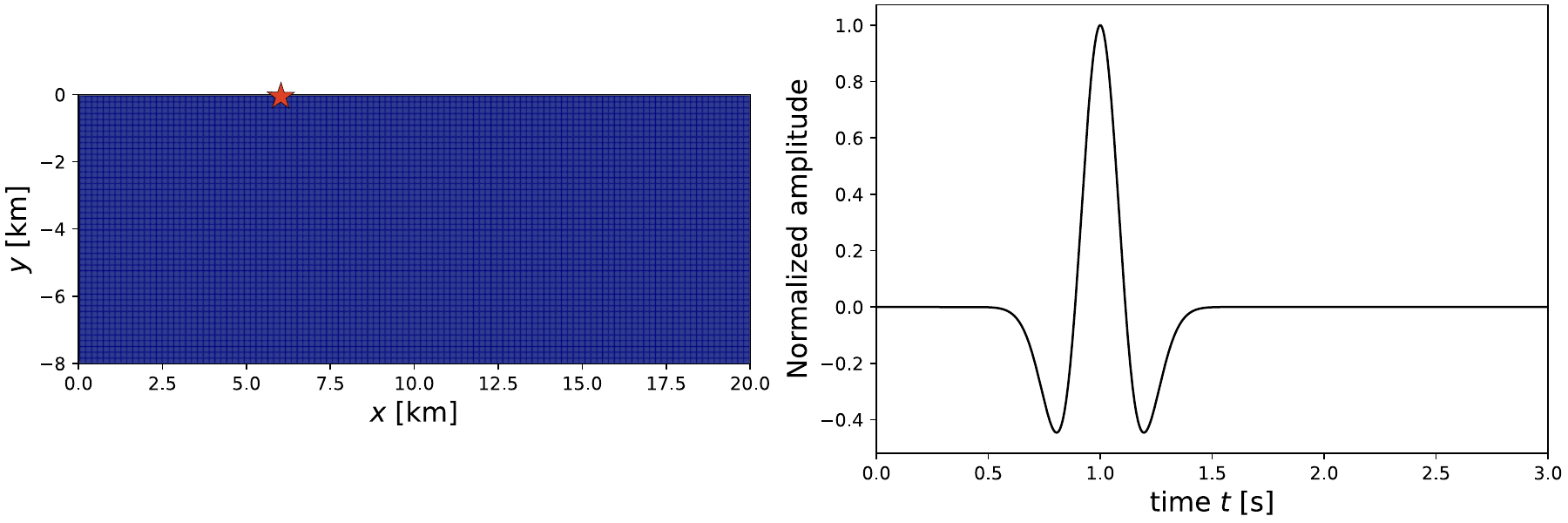}
    \caption{Model geometry and source configuration. The meshed computational domain is shown in the left panel, where the red star denotes the seismic source. The right panel displays the $\SI{2}{Hz}$ Ricker wavelet used as the source time function.}
    \label{fig:simulation_setting1}
\end{figure}

The single source is located on the free surface at the coordinate $(\SI{6}{km}, 0)$ and is implemented as an upward-pointing single force with a central frequency of $\SI{2}{Hz}$ using a Ricker wavelet (see Fig.~\ref{fig:simulation_setting1}). The source amplitude is scaled such that the resulting ground-motion levels fall between low and high seismic-noise conditions, consistent with established terrestrial seismic-noise models~\cite{peterson1993observations}.

\begin{figure}[h]
    \centering
    \includegraphics[width=\linewidth]{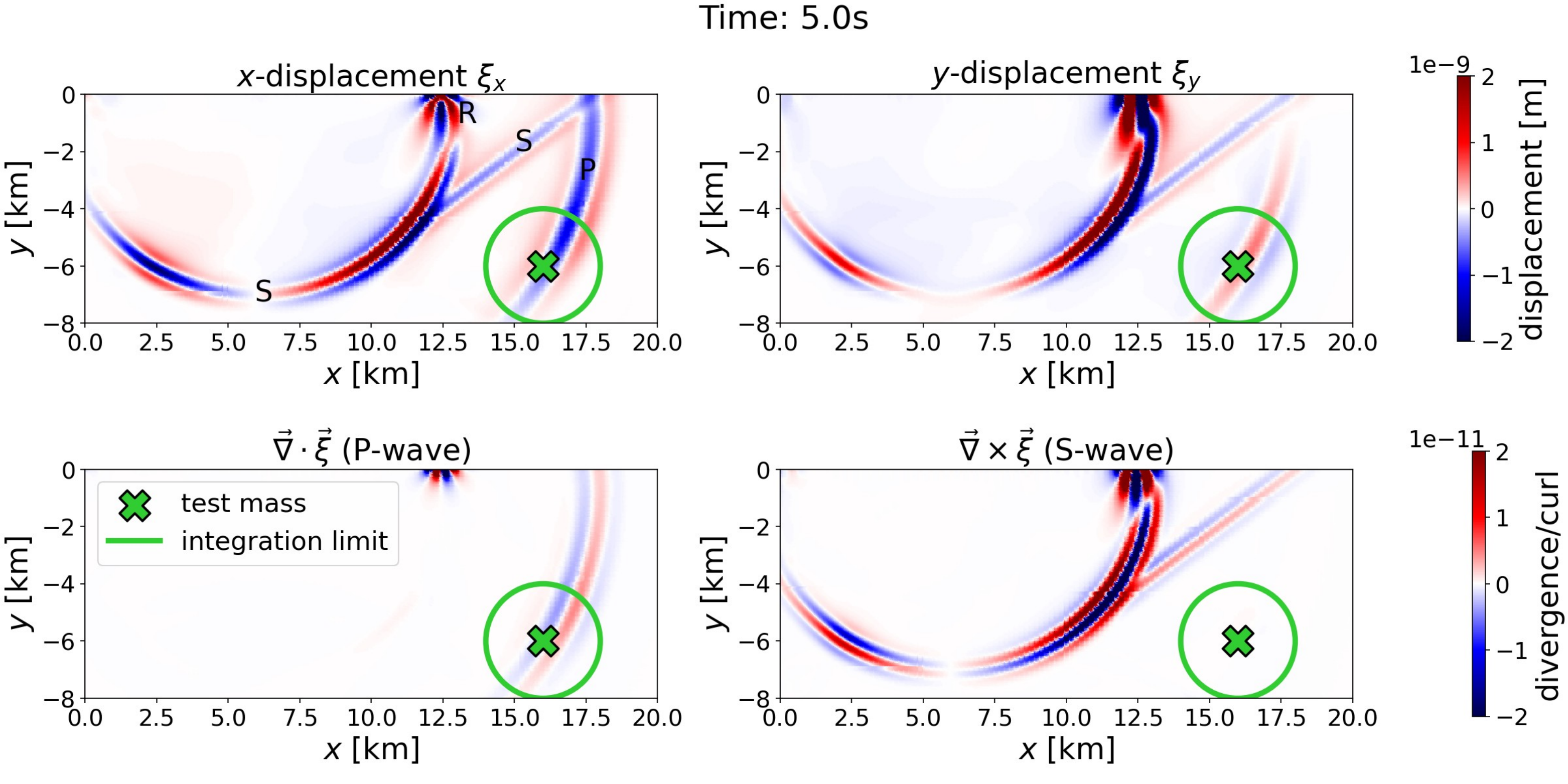}
    \caption{The displacement field and its derivatives from the numerical simulation of a single source pulse at the surface of the homogeneous medium. It shows the $x$-component of the displacement $\xi_x$ (upper left), the $y$-component $\xi_y$ (upper right), the divergence $\vec \nabla\cdot\vec \xi$ (lower left) and the curl $\vec\nabla\times\vec \xi$ (lower right) in the simulated two-dimensional domain. The green cross marks the test-mass position and the circle around it the integration domain. In the upper left plot, we have marked the different wave-field components: The P-wave (P), the S-waves (S) and the Rayleigh wave (R).}
    \label{fig:showcase_homogeneous}
\end{figure}

In Fig.~\ref{fig:showcase_homogeneous}, the wave field at time $t=\SI{5}{\s}$ after the pulse is displayed. The upper two plots show the $x$- and $y$- component of the displacement field, while the lower plots show its divergence and curl. A green cross marks the test mass that is positioned at $x=\SI{16}{\km}$ and $y=\SI{-6}{\km}$, and the main P- and S-wave energy arrives within the time windows of approximately $[4.5\,\mathrm{s},\,5.5\,\mathrm{s}]$ and $[6.5\,\mathrm{s},\,8.0\,\mathrm{s}]$, respectively. In the homogeneous case studied here, the test mass is positioned much deeper underground than actually planned for ET, so that the wave field is indeed dominated by body waves. In realistic ET geologies, body waves already dominate at much shallower depths~\cite{SiteSelectionCriteria}. The Newtonian noise at these coordinates is calculated from the wave field inside the green circle (integration domain).

Different wave fronts are visible in the displacement (see upper left plot of Fig.~\ref{fig:showcase_homogeneous}). The fastest wave front, which at this point in time has already arrived at the test-mass location, is the P-wave. As it is longitudinal, it appears in the divergence, but not in the curl. The inner circular wave is the slower S-wave. There is another (straight) S-wave branch from the interaction of the P-wave at the surface that is above the test-mass location. These two are transversal and thus are visible in the curl but not in the divergence. The last visible element has both transversal and longitudinal elements, which is the Rayleigh wave that travels along the surface at a speed similar to that of the S-wave.

At the test mass, only the body waves (P- and the S-wave fronts) lead to Newtonian noise, as the test mass is deep underground. Furthermore, it is far away from the source, so that the wave fronts appear plane. The P-wave reaches the integration domain, the green circle, at about $t=\SI{4.5}{\s}$ and leaves it again before the S-wave enters at about $t=\SI{6.5}{\s}$.

\begin{figure}[p]
    \centering
    \includegraphics[width=\linewidth]{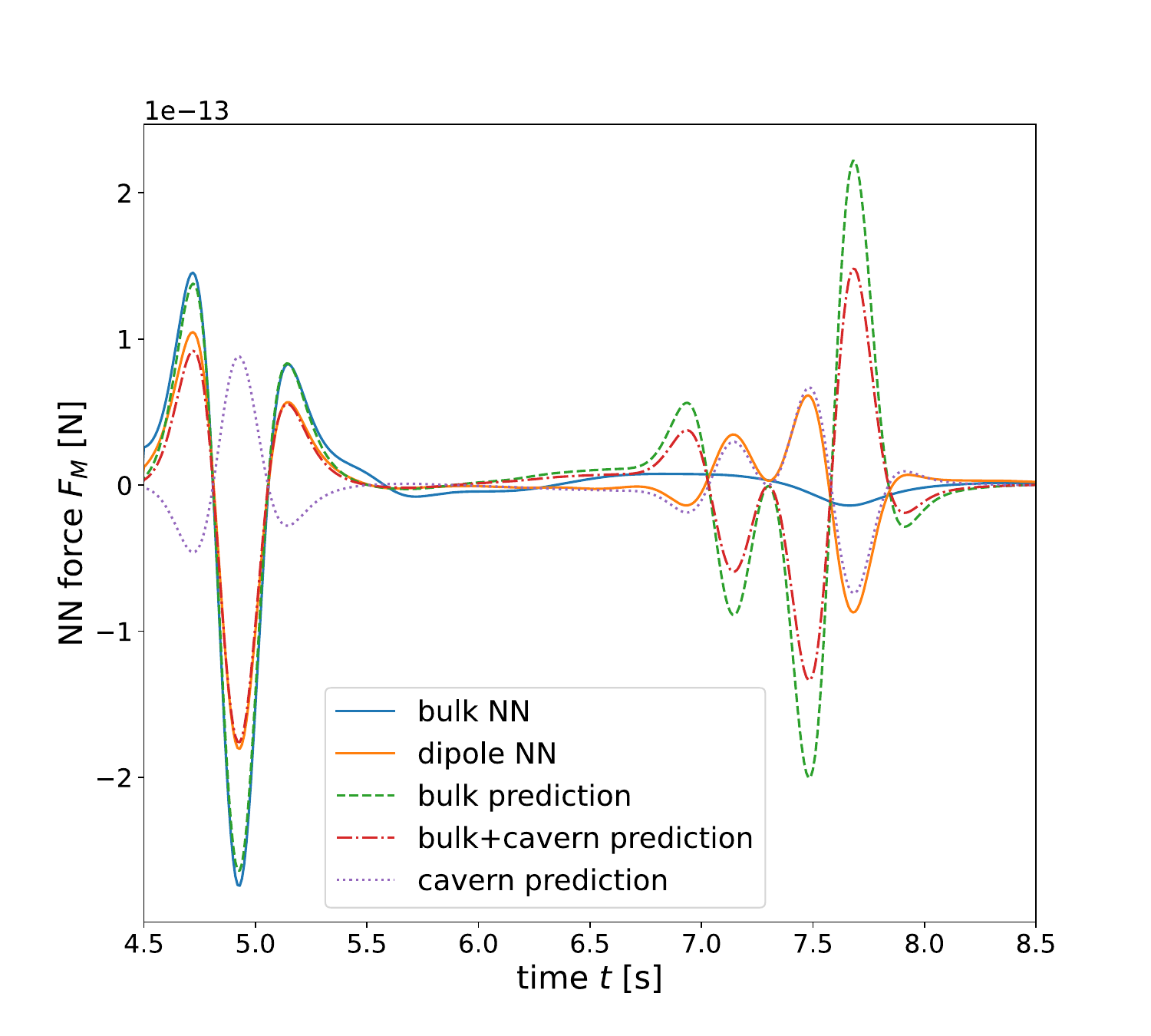}
    \caption{Different Newtonian-noise (NN) estimates calculated at the test-mass position for a single source at the surface: Shown are the result of the volume integral over density fluctuations  according to Eq.~\eqref{eq:bulk_integral} (blue, solid), and the full Newtonian-noise estimate assuming a virtual (not simulated) small spherical cavern according to the dipole formula~\eqref{eq:dipole} (orange, solid). Also shown are the analytically predicted Newtonian noise from the (virtual) displacement at the test-mass position for monochromatic plane waves: for density fluctuations only (green, dashed), for density fluctuations and cavern wall movements (= P wave, red, dashdotted), as well as cavern wall movements only (= S-wave, purple, dotted)). The left part of the plot (up to $t=\SI{6.5}{s}$) contains an almost pure plane P-wave and the right part an almost pure plane S-wave.}
    \label{fig:NN_homogeneous}
\end{figure}

Fig.~\ref{fig:NN_homogeneous} shows the Newtonian noise at the test mass as a function of time, as estimated from the simulation. Integrations are performed over the corresponding quantity within the green‑marked circle boundary of $\SI{2}{\km}$. The missing volume for the $z$-direction was simulated by assuming translation invariance in this direction to extend the divergence to three dimensions. A small spherical cavern with a radius of $\SI{5}{\m}$ was added to the density $\rho(\vec x)$ to ensure convergence. As it is small compared to all relevant wavelengths, it does not have a significant influence on the development of the seismic wave field, but it has important effects on Newtonian noise. A test mass with $M=\SI{211}{\kg}$ is assumed. 

The wave that arrives first is the P-wave. The second peaks (beyond $t = \SI{6.5}{s}$) belong to the two branches of S-waves.

One can see that the bulk integral (solid blue) only gives a contribution for the P-wave, which causes density fluctuations. It fits very well with the analytical prediction of the bulk integral result for a monochromatic, plane P-wave (dashed green). As any plane wave can be expressed as a superposition of plane monochromatic waves, this is to be expected in a homogeneous medium. For $t>\SI{6.5}{\s}$, the curves do not agree because the displacement at the test-mass position is caused by the S-wave and not by the P-wave.
The dipole integral (solid orange) has a lower amplitude than the bulk integral, as it includes a virtual cavern around the test mass, which introduces a Newtonian-noise effect with opposite direction to the bulk. However, the cavern results in a Newtonian-noise contribution also for the S-wave. When there is a P-wave ($t < \SI{6.5}{s}$), the combined prediction of bulk and cavern surface (dash-dotted red) fit well to the dipole integral. In case of an S-wave ($t > \SI{6.5}{s}$), only the cavern term (dotted purple) is relevant, which also fits well to the dipole integral.

These results validate our workflow and confirm that we can recover the analytical results in this specialized scenario. We can now move on to more interesting cases.

\subsection{An array of 30 sources in a homogeneous medium}

\begin{figure}[htbp]
    \centering
    \includegraphics[width=\linewidth]{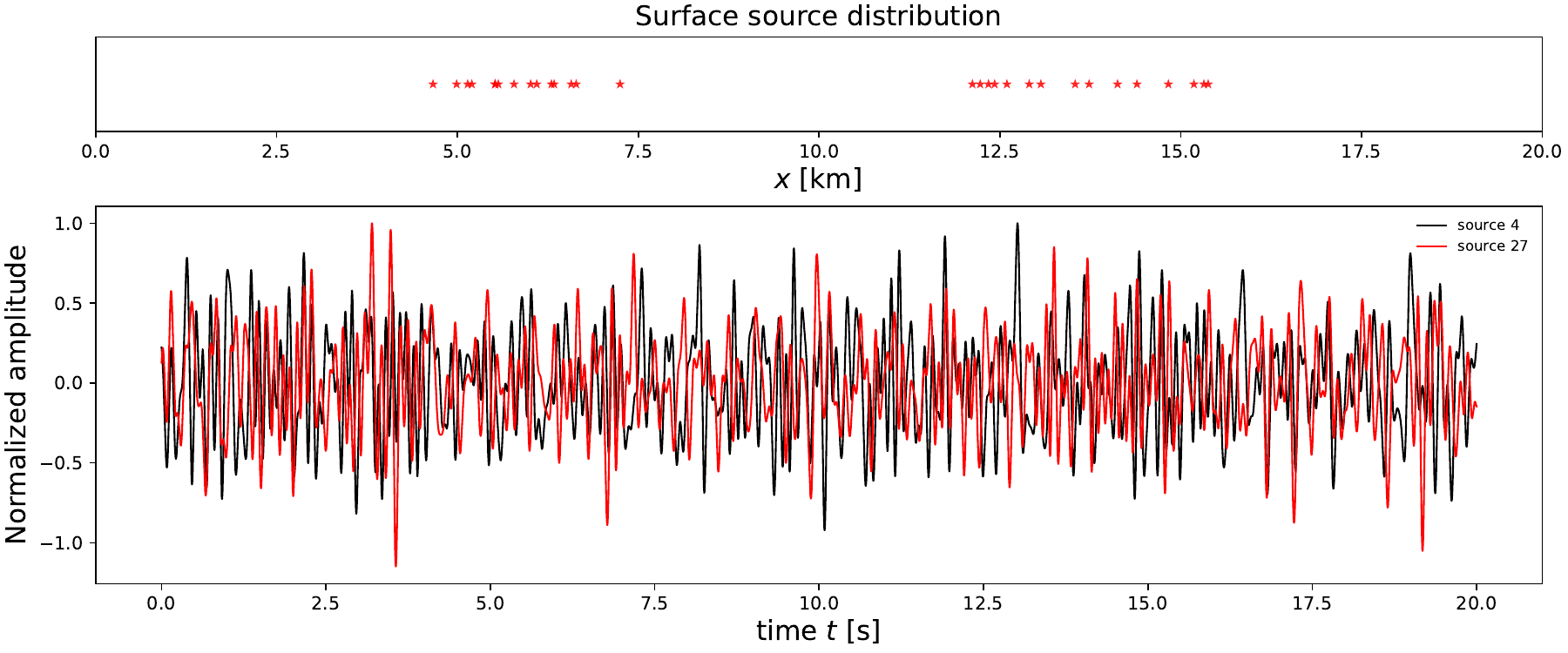}
    \caption{Source configuration for the ambient-noise simulations. The top panel shows the spatial distribution of 30 force sources along the free surface. The bottom panel presents two representative realizations of the stochastic time-domain inputs used to approximate stationary ambient seismic excitation.}
    \label{fig:simulation_setting2}
\end{figure}

In the second case, 30 randomly distributed surface sources are employed to approximate stationary ambient seismic noise. Each source is implemented as an upward-pointing single force with an independent stochastic time-domain input, band-limited to frequencies up to 10~Hz. This configuration is designed to mimic the spatially extended and temporally incoherent nature of real ambient seismic excitation (Fig.~\ref{fig:simulation_setting2}).

\begin{figure}[p]
    \centering
    \includegraphics[width=\linewidth]{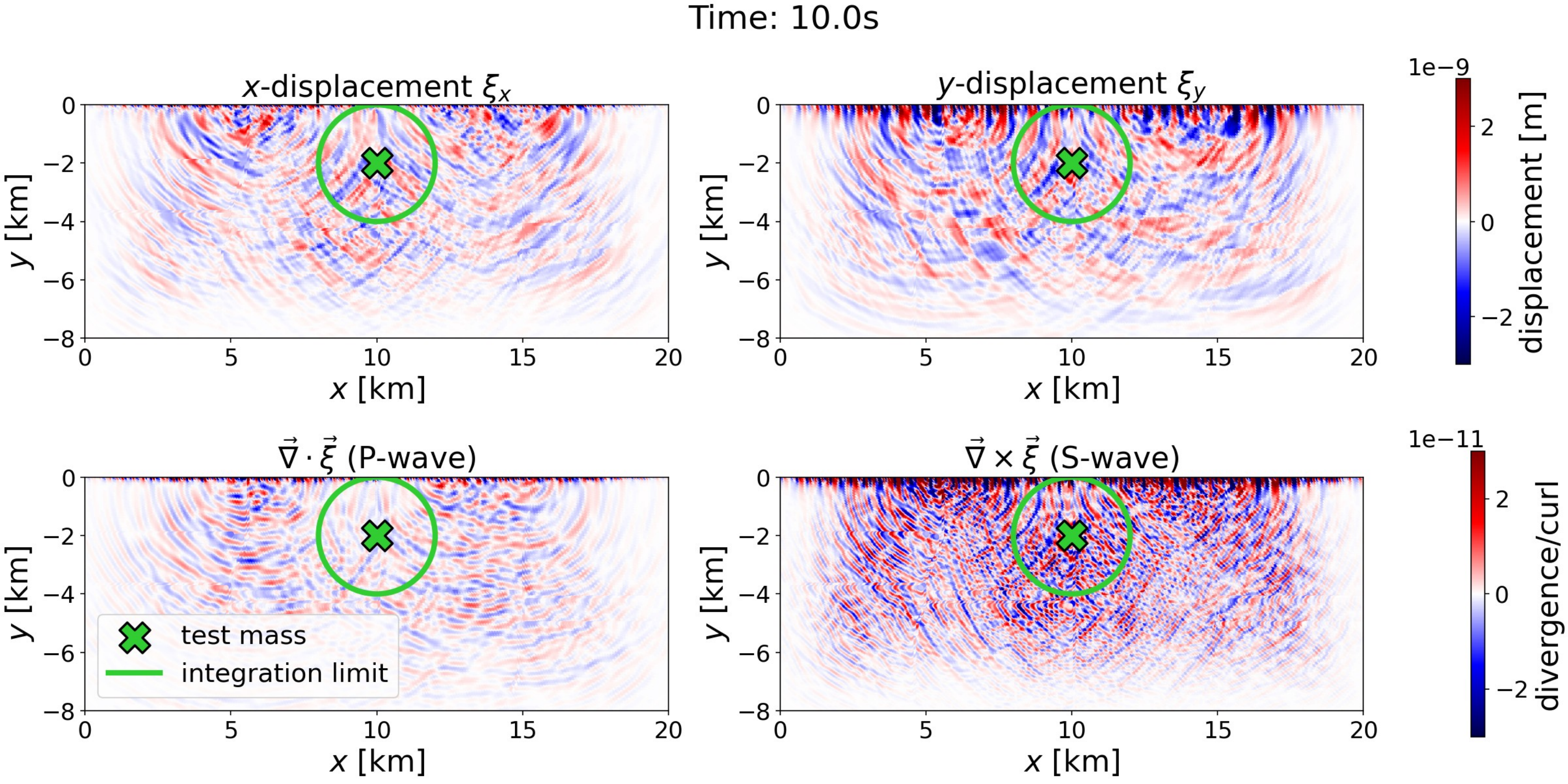}
    \caption{
    The displacement field and its derivatives from the numerical simulation of 30 randomly positioned sources at the surface of a homogeneous medium. It shows the $x$-component of the displacement $\xi_x$ (upper left), the $y$-component $\xi_y$ (upper right), the divergence $\vec \nabla\cdot\vec \xi$ (lower left) and the curl $\vec\nabla\times\vec \xi$ (lower right) in the simuated two-dimensional domain. The green cross marks the test-mass position and the circle around it the integration domain.}
    \label{fig:showcase_30sources}

\vspace{0.33cm}

\makebox[\textwidth]{
    \centering    \subfloat{\includegraphics[width=.5\linewidth]{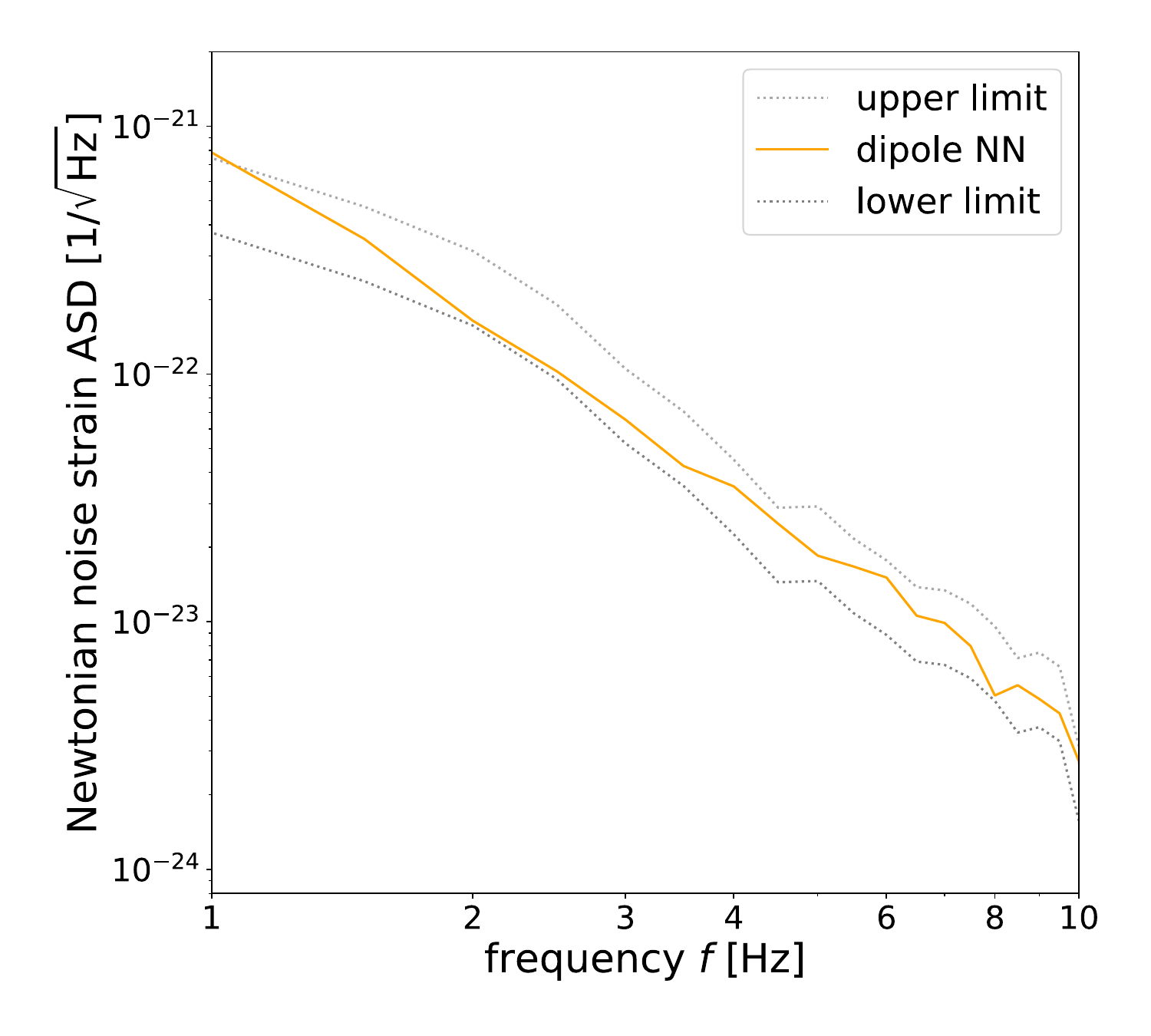}}
    \subfloat{\includegraphics[width=.54\linewidth]{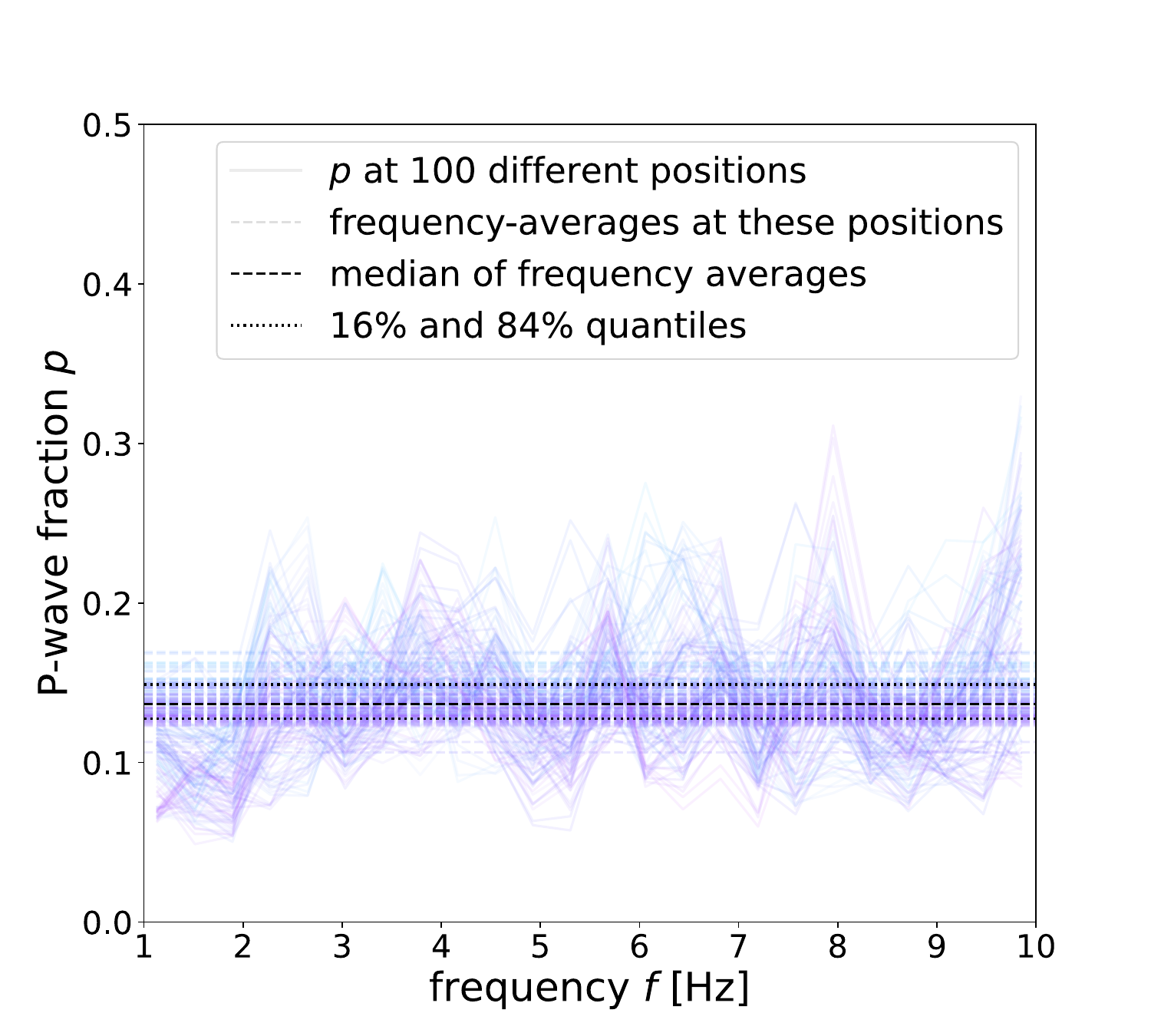}}
    }
    \caption{Left: The Newtonian-noise amplitude spectral density (ASD) at the test-mass location (see Fig.~\ref{fig:showcase_30sources}), calculated with the dipole formula (orange). Also shown are lower and upper boundaries (grey) that are estimated from the (virtual) displacement at the test-mass position, where the upper bound assumes a pure P-wave contribution and the lower bound assumes a pure S-wave contribution. Right: An estimate of the P-wave fraction $p$ from the divergence and curl in the vicinity of the test-mass position. We show $p$ at 100 points within $x\in[9.5,10.5]~\mathrm{km}$ and $y\in[-2.5,-1.5]~\mathrm{km}$ as a function of frequency (light colours in the background), as well as the frequency-averaged means at these points (light horizontal lines in the background). In addition, the black horizontal lines in the foreground show the median and the 16\% and 84\% quantiles of the frequency-averaged means.}
    \label{fig:NN_p_30sources}
\end{figure}

Fig.~\ref{fig:showcase_30sources} shows the resulting displacement field at $t=\SI{10}{s}$. The test mass is positioned right in the center and $\SI{2}{km}$ underground, so that the integration domain does not intersect with the surface. At the edges, the strength of the displacement decreases due to the implemented absorbing boundary conditions.
The higher amplitudes in the curl compared to the divergence suggest that there is a stronger S-wave content compared to P-waves.

In the left plot of Fig.~\ref{fig:NN_p_30sources}, the amplitude spectral density (ASD) of 20 seconds of Newtonian noise at the test-mass position is displayed as a function of frequency $f$.
The orange curve is, again, the dipole integral, which was scaled under the assumption of uncorrelated, free hanging mirrors with $\frac{1}{2\pi^2f^2LM}$ to obtain the strain ASD from the force ASD, where $L=\SI{10}{\km}$ is the arm length of ET. The grey curves (called \say{upper/lower limit} in the figure) represent the result under the assumption that the displacement field at the mirror is fully composed of P/S-waves (corresponding to the red/purple curve of Fig.~\ref{fig:NN_homogeneous}).
The inserted seismic spectrum is flat and scaled to realistic seismic activity ($\sim~10^{-10}\,\text{m}/\sqrt{\text{Hz}}$). Hence, the Newtonian-noise ASD falls with $f^{-2}$.
The Newtonian noise estimated with the dipole formula is, for the case of 30 sources in a homogeneous medium, constrained within the upper and lower limits but fluctuates in the full range between the boundaries. This may be due to correlations of P- and S-waves that, in contrast to the assumptions made in previous analytical estimates, originate from the same close-by sources. 

To obtain a better understanding of the actual P-wave content of the wave field, we use its divergence and curl, similar to the approach of Ref.~\cite{seismologyP}. The power spectral density (PSD) of the total wave field is given by (all quantities should be understood in frequency space)
\begin{equation}
    \EW{\xi_\text{tot}^*\xi_\text{tot}}\equiv\EW{\xi_x^*\xi_x}+\EW{\xi_y^*\xi_y}=\bracket{\frac{c_P}{2\pi f}}^2\EW{(\vec \nabla\cdot\vec\xi)^*(\vec\nabla\cdot\vec\xi)}+\bracket{\frac{c_S}{2\pi f}}^2\EW{(\vec \nabla\times\vec\xi)^*(\vec\nabla\times\vec\xi)} \, ,
\end{equation}
where the curl of a two-dimensional vector field is a scalar quantity and the derivative in frequency space leads to a factor of $k_{P,S}=\frac{2\pi f}{c_{P,S}}$, where $k$ is the wave number.
As the divergence in the body of a medium is purely influenced by P-wave displacement and the curl purely by S-wave displacement, the P-wave fraction $p$ can be expressed as
\begin{equation}
    p\equiv\frac{\EW{\xi_P^*\xi_P}}{\EW{\xi_\text{tot}^*\xi_\text{tot}}}=\frac{\EW{(\vec \nabla\cdot\vec\xi)^*(\vec\nabla\cdot\vec\xi)}}{\EW{(\vec \nabla\cdot\vec\xi)^*(\vec\nabla\cdot\vec\xi)}+\frac{c_S^2}{c_P^2}\EW{(\vec \nabla\times\vec\xi)^*(\vec\nabla\times\vec\xi)}} \, .
\end{equation}

The results of this calculation are shown on the right side of Fig.~\ref{fig:NN_p_30sources} as a function of frequency. 
To study the variations in $p$, we estimated $p$ for 100 points in the mesh in the vicinity of the test mass. 
The median of the frequency averages reads $p=0.137\pm0.02$. This is lower than the value of $p=1/3$ that has often been assumed in analytical estimates \cite{terrestialGravityFluctuations,lowerLimit}. 
If a low value of $p$ was confirmed in more realistic scenarios or by measurement, this would indicate that the estimated Newtonian-noise level is lower than assumed so far, and that the potential for its mitigation is higher than previously anticipated, as a major challenge in Newtonian-noise mitigation is the need to
properly account for the different effects of P- and S-wave components~\cite{FrancescaJointMirrorOptimization}.
Note that the actual value of $p$ depends, amongst others, on the source position and its distance from the test mass, the assumptions of the source radiation, and the geology.
Changing these parameters, we were able to produce P-wave fractions in the full range between 0 and 1, sometimes being strongly frequency dependent. We can neither exclude that they could behave differently in 3 dimensions, nor that they could converge to a given constant value in a multi-scattering regime. 
This should be investigated in more detail in the future, for example, in a more realistic three-dimensional, inhomogeneous simulation that hence also includes effects from reflections and mode conversions.

\section{Conclusions}
\label{sec:conclusions}
Making a first step towards a Newtonian-noise estimate in a real geology, in this paper, we developed a seismic wave-field simulation. We found that in the simple case of a homogeneous medium far away from the seismic source, one can very well recover the analytical Newtonian-noise models.
Furthermore, we analyzed a more complex seismic wave field with 30 sources on the surface. 
We calculated the amplitude spectral density of Newtonian noise and found that it is comparable to current analytically motivated estimates.
Moreover, we estimated the P-wave fraction. In such a scenario we found that it is lower than usually assumed, implying better prospects for Newtonian-noise mitigation. We find hints that this may be different in more realistic geologies and only conclude that the P-wave fraction should be studied with more care in the future.

Building on this proof-of-principle study, the next step is to move away from a homogeneous space to include effects of reflections and mode conversions. The final goal must be to recreate the Einstein Telescope's site geologies and obtain the most accurate Newtonian-noise estimate possible. These results could, in the future, also be used to estimate mitigation prospects and may also be used to optimize a seismic array for Newtonian-noise predictions.

\section*{Availability of Code}
The code used to produce the results of this paper is public at \url{https://github.com/lc316353/A-numerical-framework-for-Newtonian-noise-estimation}.

\section*{Acknowledgements}
We thank Jan Harms for valuable comments on the manuscript. This research was supported by the German Federal Ministry of Research, Technology and Space (BMFTR) via project 05A2023 under grant number 05A23PA1. Shi Yao was partly funded through the STARK Program (grant number 46SK0336X) by the Federal Ministry of Economic Affairs and Energy (BMWE).
\printbibliography
\end{document}